\documentclass[journal]{IEEEtran}
\usepackage{graphicx}
\usepackage{multirow}
\usepackage{array}
\usepackage{url}
\usepackage{amsmath}
\usepackage{authblk}
\usepackage{NotationMacros}

\ifCLASSINFOpdf
\else
\fi

\begin{document}

\title{Perceptual Evaluation of Extrapolated Spatial Room Impulse Responses From a Mono Source}

\author{
    \IEEEauthorblockN{Ben Heritage\IEEEauthorrefmark{1}, Fiona Ryder\IEEEauthorrefmark{1}}, 
    \IEEEauthorblockN{Michael McLoughlin\IEEEauthorrefmark{2}, Karolina Prawda\IEEEauthorrefmark{2}}
    \thanks{This research was supported by UKRI Innovate UK grant no. 10123863. \newline
    \IEEEauthorrefmark{1}Bonza Music LTD, London, UK\newline
    \IEEEauthorrefmark{2}AudioLab, School of Physics, Engineering and Technology, University of York, York, UK}
}

\markboth{}%
{}

\maketitle

\begin{abstract}

Immersion in virtual and augmented reality solutions is reliant on plausible spatial audio. However, plausibly representing a space for immersive audio often requires many individual acoustic measurements of source-microphone pairs with specialist spatial microphones, making the procedure time-consuming and expensive. In this study, we evaluate the plausibility of extrapolated and spatialised Room Impulse Responses (RIRs) by using a 3-Alternative Forced Choice (3AFC) listening test. The stimuli comprised of RIRs from three spaces convolved with speech, orchestral, and instrumental music. When asked to select which stimuli was artificial out of one extrapolated and two real stimuli, an overall accuracy of $\boldsymbol{38\%}$ was achieved from 20 participants ($\boldsymbol{5}$ percentage points above the expected guessing rate). Given the listening test result, this study shows that it is possible to extrapolate plausible spatial RIRs from mono measurements, decreasing the need for time and specialist equipment in acoustic measurements.

\end{abstract}


\IEEEpeerreviewmaketitle

\section{Introduction}\label{section:Introduction}

\IEEEPARstart{I}{n} immersive audio applications, user immersion greatly benefits from auditory cues for localising the sound source and the listener's own position~\cite{warp2022moved}. To accurately replicate the space's acoustic properties digitally, an acoustic measurement or Room Impulse Response (RIR) is required. Within a RIR, direction of arrival (DoA) of the direct sound carries the cues for the source location, while DoA and timbre of the early reflections specify the user's position~\cite{wallach1949precedence}. Mono RIRs do not contain directional information, and do not give a sense of directional or spatial information. However, multi-channel RIRs, such as Spatial RIRs (SRIRs), contain information regarding the DoA and timbre of the direct sound and the early reflections. Converting SRIRs to a stereo file in the form of Binaural RIRs (BRIRs), allows the user to experience the spatial aspects of an SRIR whilst using headphones. Using BRIRs can give the end-user a sense of source location and spatial awareness; additionally the immersion can be enhanced by using head-tracking and updating the BRIR in-use accordingly~\cite{potter2022relative}.

However, plausible acoustic representation of distinct places in immersive applications typically require a large database of SRIRs (or BRIRs). Within each individual space, a variety of placements and orientations are desired in order to capture the acoustic properties thoroughly and allow for convincing interpolation between measurements~\cite{qiao2024multi}. On top of this, many commercial immersive audio applications call for a variety of acoustically different spaces in order to provide users with an interesting auditory experience.

Such databases are time consuming to attain and require capturing many source-microphone positions with expensive dedicated equipment such as the KU-100~\cite{KU-100} or a soundfield microphone. Additionally, environments with intermittent noise and limited availability or access are challenging when collecting measurements. Thus, measurements can be rushed, take longer than expected, or result in a lower quality recording.

Mono RIRs are readily available online, and when recording bespoke measurements, the time doing field recording is significantly reduced and is comparatively inexpensive. On top of this, it is possible to extrapolate and spatialise mono RIRs to SRIRs, avoiding the problems associated with manual SRIR acquisition. This study presents a perceptual analysis of spatially extrapolated SRIRs from a single mono RIR source, resulting in a reduction in time and complexity, which is beneficial in collating a diverse database.

Within this paper, section \ref{section:Listening Test Methodology} details the listening test design, section \ref{section:Results} shares the results and analysis of the listening test, and finally section \ref{section:Conclusion} concludes the findings and discuss the limitations of our presented approach.

\section{Listening Test Methodology}\label{section:Listening Test Methodology}

The listening test has been designed on the use-case of a mixed database comprised of both real and extrapolated SRIRs. In which, the user can alternate between distinct acoustic spaces at will, depending on the audio context (eg. speech, choral, instrumental). Following research on the effectiveness of listening test paradigms~\cite{AuditoryIllusions} of authenticity, transfer-plausibility, and plausibility within immersive audio, the 3-Alternative Forced Choice (3AFC) test was chosen. This transfer-plausibility paradigm detailed was chosen as the authenticity paradigm was shown to be too strict for the purpose of immersive applications~\cite{brinkmann2017authenticity}. Furthermore, the plausibility paradigm is shown to be less efficient and is also less effective when comparing multiple test conditions~\cite{AuditoryIllusions}. An intermediary test (between plausibility and transfer-plausibility) 2AFC was suggested in~\cite{AuditoryIllusions}, however, the 3AFC paradigm gave a higher sensitivity. Additionally to these points, the 3AFC transfer-plausibility paradigm aligns closely with the use-case.

The test stimuli were generated using dry recordings of speech, orchestral and instrumental samples, convolved with extrapolated and real BRIRs (generated from SRIRs). The speech samples contain a female child, a female adult, and a male adult reading sentences 1-5 from the Harvard list 2 obtained from the TSP Speech Database~\cite{HarvardSentences}~\cite{kabal2002tsp}. Orchestral samples were collected on OpenAIR~\cite{OpenAir} from the recordings 2, 5, 14, and 15. The instrument samples are of isolated drums, guitar~\cite{Avad-VR}, and piano, which were chosen for their percussive and diverse tonal and rhythmic properties. The piano samples were created using a midi archive of classical music~\cite{midi-kunstlerfuge}, where a solo piano midi file was played through a dry piano sample from PianoBook~\cite{pianobook}. The orchestral and instrumental samples were all trimmed to three 3-second samples from each source. Each stimuli had a 10ms fade-in and fade-out to mitigate audible clicks in the audio.

\begin{figure}[!h]
    \begin{center}
    \includegraphics[width=8.8cm]{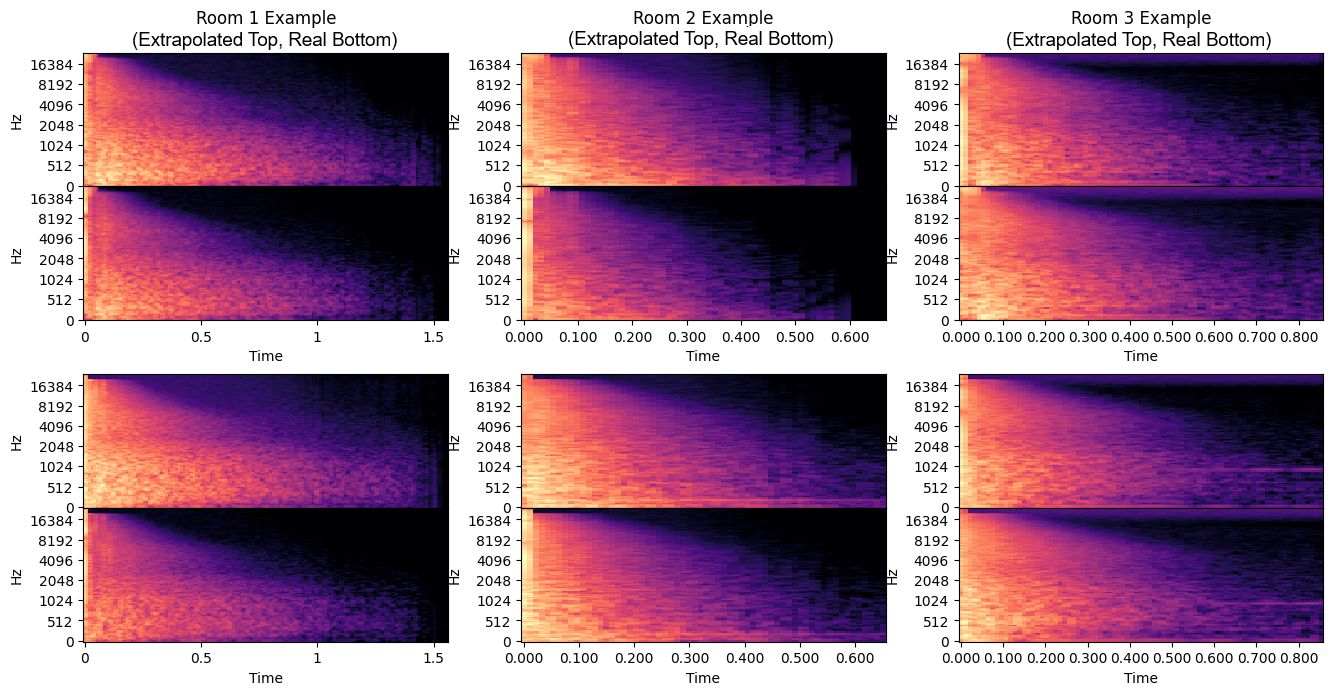}
    \caption{Example spectrograms of the real and extrapolated BRIRs for each of the three test rooms (at the same source-microphone locations)}
    \label{example_spectrogram}
    \end{center}
\end{figure}

The three spaces chosen for the listening test had differing $T_{60}$ as shown in Table \ref{rt60_rooms} (analysed using DecayFitNet~\cite{decayfitnet}). The real-world SRIRs were recorded using the spcmic, a 3rd-order ambisonic spatial microphone using 84 high-quality MEMS microphones~\cite{spcmic}. These recordings are of three sources and three microphone placements whose relative distances and orientations are the same across all real and extrapolated recording locations. This resulted in 5 distinct source-microphone distances and 9 different azimuth angles. Using the mono channel of a reference spcmic measurement, the extrapolated SRIRs were generated using Bonza Music's in-house algorithm, giving an output of 7th-order ambisonic SRIRs in the same source-microphone pairings as their real-world counterparts. Each of the real and extrapolated SRIRs was converted to BRIRs using the SADIE-II HRIRs. The direct sounds of all BRIRs were replaced with the SADIE-II BRIRs given the correct DoA and power matched to the original direct sound, preserving the Direct-to-Reverberant Ratio (DRR)~\cite{sadie}. This was implemented to entice listening test participants to make decisions based on the spatial impression of the entire BRIRs rather than focusing on differences between the direct sounds. 
Figure \ref{example_spectrogram} shows example spectrograms for each analyzed room, showcasing the frequency differences between the extrapolated and real BRIRs.

\begin{table}[!h]
    \begin{center}
    \caption{$T_{60}$ per octave and wideband (WB) for the test rooms}
    \label{rt60_rooms}
    \begin{tabular}{|p{0.7cm}|p{0.45cm}|p{0.45cm}|p{0.45cm}|p{0.45cm}|p{0.45cm}|p{0.45cm}|p{0.45cm}|p{0.9cm}|}
    \hline
    \multirow{2}{*}{\textbf{Room}} & \multicolumn{7}{c|}{\bm{$T_{60}$}\textbf{~(s)} \textbf{per Octave Band (Hz)}} & \multirow{2}{0.7cm}{\bm{$T_{60}$}\textbf{~(s) WB}} \\
    \cline{2-8}
    & \textbf{125} & \textbf{250} & \textbf{500} & \textbf{1k} & \textbf{2k} & \textbf{4k} & \textbf{8k} & \\
    \hline
    1 & 1.31 & 1.54 & 1.68 & 1.52 & 1.33 & 0.89 & 0.61 & 1.38\\ %Hes Church
    \hline
    2 & 0.80 & 0.74 & 0.65 & 0.60 & 0.57 & 0.47 & 0.33 & 0.80\\ %Voodoo Daddys
    \hline
    3 & 0.64 & 0.61 & 0.60 & 0.66 & 0.61 & 0.59 & 0.59 & 0.68\\ %Hope Ruin
    \hline
    \end{tabular}
    \end{center}
\end{table}

After convolving the dry stimuli with each of the BRIRs, the maximum value of each stimulus was normalised to -\(6\)\,dB. Since the longest WB $T_{60}$ was 1.38\,s, each reverberated stimuli was trimmed to the length of the dry sample plus 1.5\,s to make sure that the reverberation tail could fully decay and could be heard in isolation for each sound. Additionally, every stimuli had a 200\,ms fade-out as well to prevent audible clicks.

\begin{figure}[!h]
    \begin{center}
    \includegraphics[width=7cm]{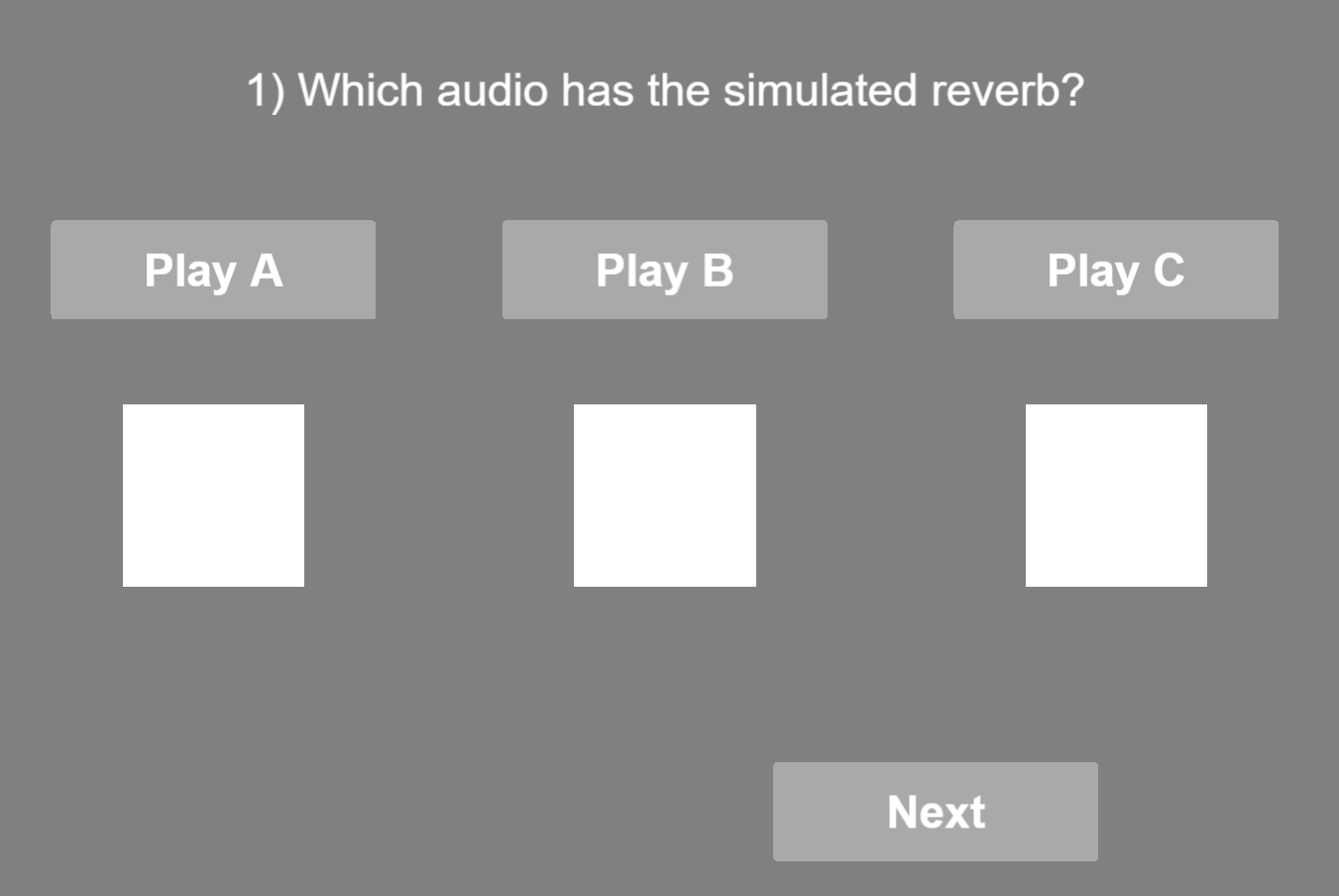}
    \caption{A snapshot of the 3AFC paradigm in the listening test, created in PsychoPy\,\cite{peirce2019psychopy2}}
    \label{3afc_screen}
    \end{center}
\end{figure}

During the test, each stimulus set contains three unique source-microphone pairs in the same room (one extrapolated and two real), and of the same test type as shown in Table \ref{stim_test_type}. The order of the questions and the location on the screen (stimuli A, B, or C) of the extrapolated sound on the screen are randomised for each participant. Figure \ref{3afc_screen} depicts the listening test screen. The participant was only allowed to progress to the next question after providing an answer. The listening test contains one of every test type in Table \ref{stim_test_type} for each of the rooms, resulting in 30 stimuli sets that are used for all participants.

\begin{table}[!h]
    \begin{center}
    \caption{Stimuli test types}
    \label{stim_test_type}
    \begin{tabular}{|p{2.4cm}|p{2.1cm}|p{2.8cm}|}
        \hline
        \textbf{Speech} & \textbf{Orchestral} & \textbf{Instrumental} \\ \hline
        \textbf{1}: Same-Speaker Same-Sentence & \textbf{1}: Same-Piece Same-Extract & \textbf{1}: Same-Instrument Same-Extract \\ \hline
        \textbf{2}: Different-Speaker Same-Sentence & \textbf{2}: Same-Piece Different-Extract & \textbf{2}: Same-Instrument Different-Extract \\ \hline
        \textbf{3}: Same-Speaker Different-Sentence & \textbf{3}: Different-Piece & \textbf{3}: Different-Instrument \\ \cline{0-0}
        \textbf{4}: Different-Speaker Different-Sentence & & \\ 
        \hline
    \end{tabular}
    \end{center}
\end{table}

The listening test was completed in person at the University of York's AudioLab with participants over the age of 18. To keep the test fair over all participants, Beyerdynamics DT990 pro~\cite{beyer} headphones were used for all tests. The participants were instructed to choose a level that was comfortable for them, without the option to alter it afterwards, but did not exceed \(80\)\,dB.

\section{Results}\label{section:Results}

In total the test was completed by 20 participants, 7 female and 13 male, aged $21-44$ with a mean ($\mu$) of $26.7$ and standard deviation ($\sigma$) of $5.4$. The participants had $0.5-20$ years of working within the audio industry ($\mu: 5.8$, $\sigma: 5.0$), with $60\%$ working within audio research. $70\%$ of participants had experience with listening tests, and no participants declared any hearing impairment.

\subsection{Qualitative Analysis}

\begin{figure}[!h]
    \begin{center}
    \includegraphics[width=9cm]{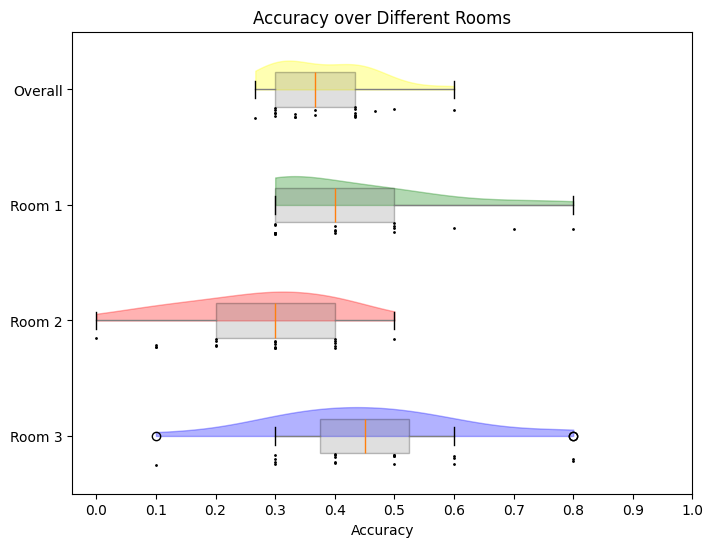}
    \caption{Raincloud plot showing the accuracies between the rooms}
    \label{Rooms_Raincloud}
    \end{center}
\end{figure}

Figure \ref{Rooms_Raincloud} depicts the distribution of each of the participant's overall accuracy in a given category, with accuracies for each room can be compared to the overall accuracy distribution. The overall accuracy for the listening test ($\mu$) was $38\%$, with $\sigma$ of $8\%$, and the median was $37\%$. It is worth noting that the age, years worked in audio, and self-prescribed confidence level of participants had no positive correlation on the test's accuracy (with Pearson Correlation Coefficients of -0.13, -0.11, and -0.06 respectively). The overall accuracy was $5$ percentage points above the guessing rate of $33\%$, proving that the task of differentiating between real and extrapolated BRIRs was challenging. From the WB $T_{60}$ values shown in Figure \ref{rt60_rooms}, we would assume that room 1 (with the largest WB $T_{60}$) should be the easiest for the participants to identify of the three rooms, due to the lower DRR allowing for a longer time to examine the late reverberation. However, the results suggest that Room 3 was the easiest to distinguish, with the shortest WB $T_{60}$, the medium length tail being the hardest to distinguish, and the longest in-between. Hence, the length of the reverb tail in this test is not correlated with the accuracy.

\begin{figure}[!h]
    \begin{center}
    \includegraphics[width=9cm]{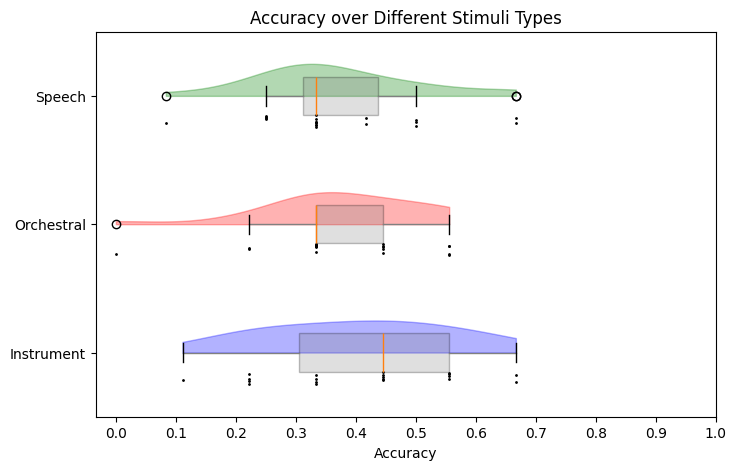}
    \caption{Raincloud plot showing the accuracies between the stimuli types}
    \label{Stimuli_Raincloud}
    \end{center}
\end{figure}

Figure \ref{Stimuli_Raincloud} shows the distributions of accuracy over the stimuli types. The speech and orchestral achieved similar accuracies at around the guessing rate, with the instrument accuracy situated higher. Due to the complexity and busyness of the orchestral stimuli, is is expected that this would have lower accuracy. As the authors expected, the instrument stimuli gave the best accuracy, a possible reason for this is the overtly rhythmic qualities compared to the other stimuli types, giving impulse-like responses. 

\begin{figure}[!h]
    \begin{center}
    \includegraphics[width=6cm]{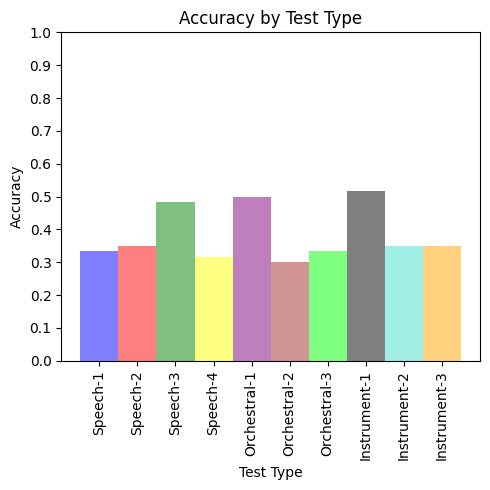}
    \caption{Average Accuracy for Each Test Type}
    \label{Stimuli_Test-type}
    \end{center}
\end{figure}

Figure \ref{Stimuli_Test-type} confirms the authors' expectation that the tests where the stimuli are identical (apart from the speech stimuli) result in higher accuracy than when stimuli differ in terms of excerpt or origin. Surprisingly, Speech-3 (Same-Speaker Different-Sentence) test type resulted in a higher accuracy than the Speech-1 (Same-Speaker Same-Sentence) test type. Although the true reason for this is currently unknown, we expect that this effect would diminish with a larger participant group.

\begin{figure}[!h]
    \begin{center}
    \includegraphics[width=9cm]{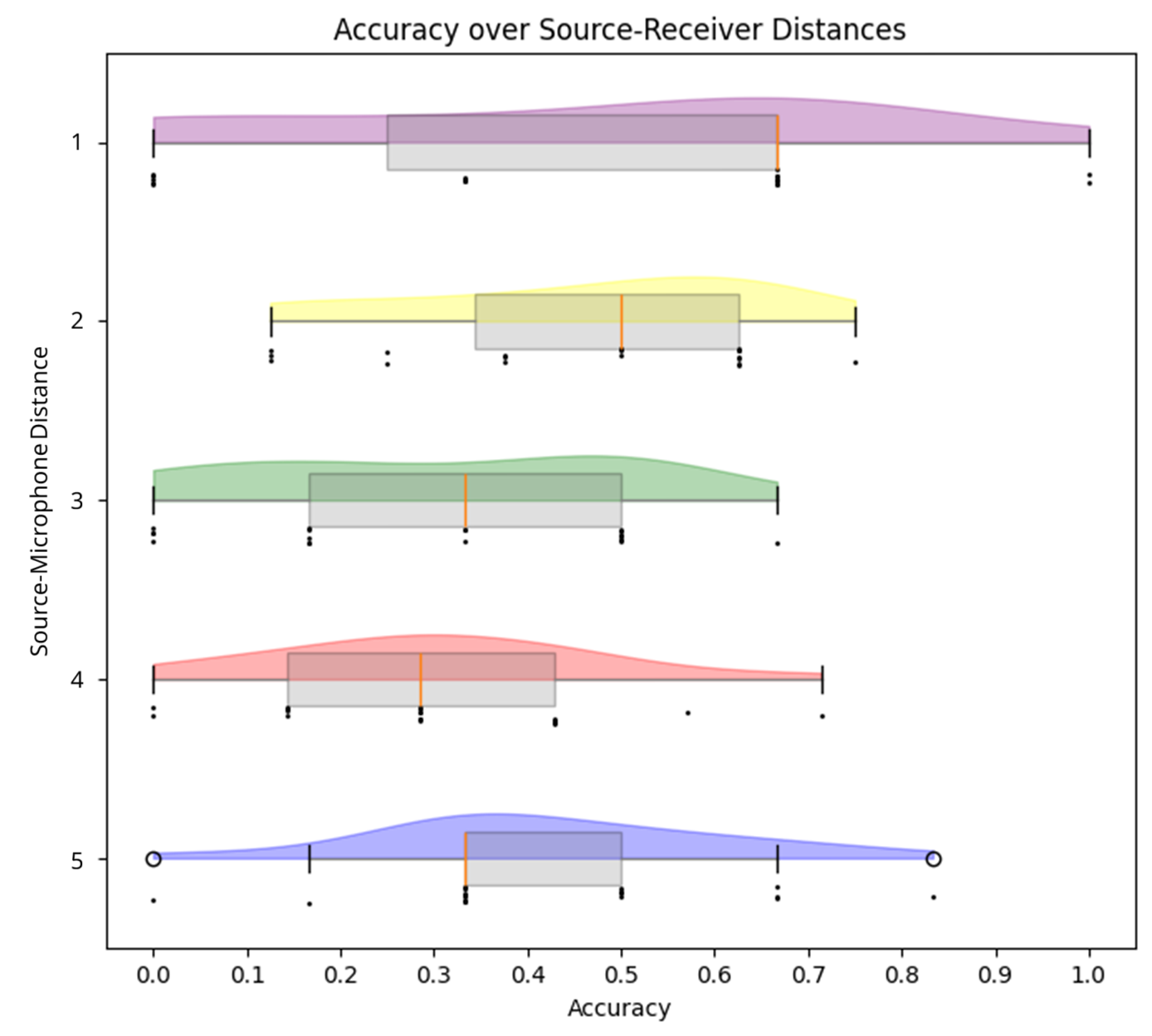}
    \caption{Raincloud plot showing the accuracies between the source-microphone distance (shortest to longest, 1-5)}
    \label{Distance_Raincloud}
    \end{center}
\end{figure}

\begin{figure}[!h]
    \begin{center}
    \includegraphics[width=4cm]{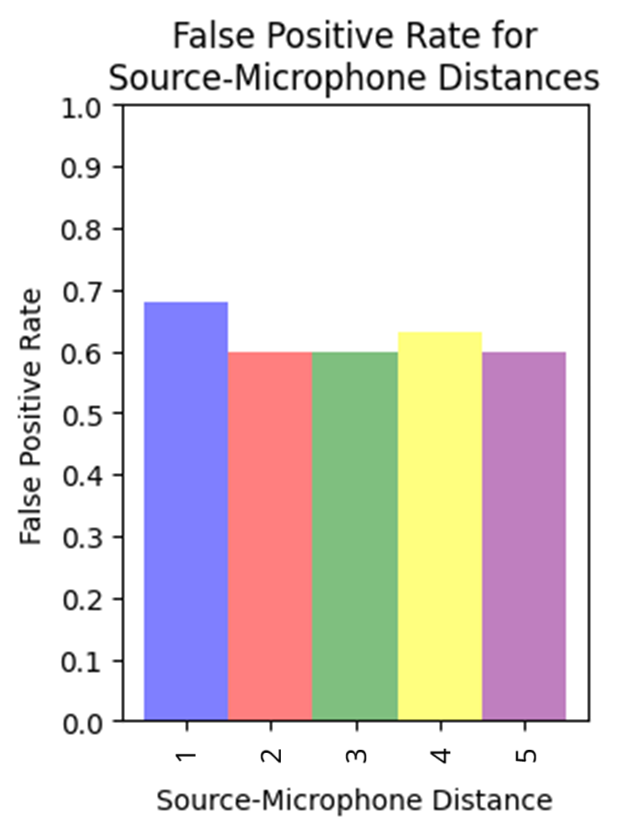}
    \caption{Source-Microphone Distance False-Positive Rate (shortest to longest, 1-5)}
    \label{Distance_False-Positive}
    \end{center}
\end{figure}

Figure \ref{Distance_Raincloud} shows the distribution of accuracies over the source-microphone distances. We would expect the larger distances to be higher in accuracy due to the lower DRR, where the user can hear more of the reverb tail before it falls below the noise floor. However the stimuli with higher DRR (distances 1-2) are more accurately identified, with a dip in the middle of the range (3-4). Figure \ref{Distance_False-Positive} illustrates the distribution of false positives for each of the source-microphone distances. They are all close to an expected false positive rate of $66.7\%$, therefore suggesting that the DRR does not affect the identification success, or does so in a small degree.

The analysis of open comments regarding participants reasoning of their choices showed that the higher scorers tended to state the spatialisation, timbre, fullness, and early reflection frequency response as the main reference for their decisions. The participants who scored closer to the guessing rate stated it was difficult to distinguish and had to rely on the reverb tail's length, frequency components, and naturalness. Hence, it can be inferred that the spatialisation of a mono RIR and the closer matching of the early reflection timbre are areas to develop further to bridge the gap between the overall acuracy and the guess rate.

\subsection{Binomial Proportion Tests}

Following the analysis on the 3AFC outcomes in ~\cite{AuditoryIllusions}, the binomial test has been completed for the whole test and sections of the test in order to assess if there are differences within each independent variable. If the spatialised mono RIRs ate indistinguishable from their real-world counterparts, the participants' answers should be correct in $33.3\%$ of the test questions, following the test guessing rate. Therefore, the null hypothesis $H_0$ for each test is that the accuracy $p$ is equal to the guessing rate of $\frac{1}{3}$ and the alternative hypothesis $H_1$ is that the accuracy is larger than the guessing rate $\frac{1}{3}$:

\[H_{0}:p=\frac{1}{3}, \>\>\>\>\> H_{1}:p>\frac{1}{3}.\]

For all binomial proportion tests, the $\alpha$ value is set to $0.05$. Table \ref{tab:binom} details all the sections of binomial tests completed on the data.

\begin{table}[!h]
    \begin{center}
    \caption{Binomial Proportion Tests for All Sections of Data}
    \label{tab:binom}
    \begin{tabular}{|p{1.2cm}|p{1.2cm}|p{0.5cm}|p{0.9cm}|p{0.5cm}|p{1cm}|}
        \hline
        \textbf{Category} & \textbf{Section} & \textbf{N} & \textbf{Critical Value} & \textbf{x} & $\boldsymbol{H_{0}}$ \\ \hline\hline

        \textbf{Overall} & & 600 & 219 & 230 & Rejected \\ \hline\hline

        \textbf{Rooms} & 1 & 200 & 78 & 84 & Rejected \\ \cline{2-6}
        & 2 & 200 & 78 & 54 & \textbf{Accepted} \\ \cline{2-6}
        & 3 & 200 & 78 & 92 & Rejected \\ \hline\hline
        
        \textbf{Stimuli} & Speech & 240 & 92 & 89 & \textbf{Accepted} \\ \cline{2-6}
        \textbf{Type} & Orchestral & 180 & 70 & 68 & \textbf{Accepted} \\ \cline{2-6}
        & Instrument & 180 & 70 & 73 & Rejected \\ \hline\hline

        \textbf{Src-Rec} & 1 & 69 & 30 & 22 & \textbf{Accepted} \\ \cline{2-6}
        \textbf{Distance} & 2 & 141 & 56 & 56 & Rejected \\ \cline{2-6}
        & 3 & 141 & 56 & 57 & Rejected \\ \cline{2-6}
        & 4 & 136 & 54 & 50 & \textbf{Accepted} \\ \cline{2-6}
        & 5 & 113 & 46 & 45 & \textbf{Accepted} \\ \hline
    \end{tabular}
    \end{center}
\end{table}

For the overall test, $H_0$ was rejected, and out of the 11 sections, $H_0$ was accepted 6 times and rejected 5 times. The sections where $H_0$ was accepted were: Room 2, Speech Stimuli, Orchestral Stimuli, and Source-Microphone distances 1, 4 and 5. For the sections where $H_0$ was accepted, the answers collected were not statistically significant from the guessing rate, and the extrapolated SRIRs in these tests are considered to be plausible.

% These binomial proportion tests highlight that with this sample size, some of the test outcomes are statistically significant to discount the hypothesis that the participants are guessing at random. A few subsections (Instrument and Source-Microphone distances 2 and 3) that were tested are only statistically significant by a small number of questions, therefore with a larger sample size, these could lead to the $H_{0}$ being accepted. This inconclusive result shows that in the context of real BRIRs, the method used to extrapolate requires further development.

\section{Conclusion}\label{section:Conclusion}

In this study, we tested the transfer-plausibility of extrapolated SRIRs from a mono source. We conducted the 3AFC listening test comparing the extrapolated SRIRs with their real-world counterparts. The results show that in more than half of the test sections, the results are not statistically significant from the guessing rate. Considering this, the method used to extrapolate mono RIRs, performs well in generating plausible SRIRs of the acoustic space, even though mono RIRs do not contain any spatial information.

Further refinements are needed for perceptual enhancement in the tests where $H_0$ was rejected. The key areas of focus are the spatialisation of the late reverberation and the the timbre of the early reflections, as suggested by the higher scoring listening test participants. With this in mind, one of the key difficulties when presented with a mono RIR, is the lack of spatial information. Hence, matching the spatialisation of the late reverberation to it's real-world counterpart is challenging and an area for further research.

\bibliographystyle{unsrt}
\bibliography{bibfile}

\begin{thebibliography}{10}

\bibitem{warp2022moved}
Richard Warp, Michael Zhu, Ivana Kiprijanovska, Jonathan Wiesler, Scot Stafford, and Ifigeneia Mavridou.
\newblock Moved by sound: How head-tracked spatial audio affects autonomic emotional state and immersion-driven auditory orienting response in {VR Environments}.
\newblock In {\em Audio Engineering Society Convention 152}. Audio Engineering Society, 2022.

\bibitem{wallach1949precedence}
Hans Wallach, Edwin~B Newman, and Mark~R Rosenzweig.
\newblock A precedence effect in sound localization.
\newblock {\em The Journal of the Acoustical Society of America}, 21(4\_Supplement):468--468, 1949.

\bibitem{potter2022relative}
Thomas Potter, Zoran Cvetkovi{\'c}, and Enzo De~Sena.
\newblock On the relative importance of visual and spatial audio rendering on {VR} immersion.
\newblock {\em Frontiers in Signal Processing}, 2:904866, 2022.

\bibitem{qiao2024multi}
Yue Qiao, Ryan~Miguel Gonzales, and Edgar Choueiri.
\newblock A multi-loudspeaker binaural room impulse response dataset with high-resolution translational and rotational head coordinates in a listening room.
\newblock {\em Frontiers in Signal Processing}, 4:1380060, 2024.

\bibitem{KU-100}
Neumann {KU} 100.
\newblock \url{https://www.neumann.com/en-us/products/microphones/ku-100}.
\newblock Accessed: 2025-04-01.

\bibitem{AuditoryIllusions}
Nils Meyer-Kahlen, Sebastian Schlecht, Sebasti{\`a} V~Amengual Gar{\'\i}, and Tapio Lokki.
\newblock Testing auditory illusions in augmented reality: Plausibility, transfer-plausibility, and authenticity.
\newblock {\em Journal of the Audio Engineering Society}, 72(11):797--812, 2024.

\bibitem{brinkmann2017authenticity}
Fabian Brinkmann, Alexander Lindau, and Stefan Weinzierl.
\newblock On the authenticity of individual dynamic binaural synthesis.
\newblock {\em The Journal of the Acoustical Society of America}, 142(4):1784--1795, 2017.

\bibitem{HarvardSentences}
Ernst~H Rothauser.
\newblock {IEEE} recommended practice for speech quality measurements.
\newblock {\em {IEEE} Transactions on Audio and Electroacoustics}, 17(3):225--246, 1969.

\bibitem{kabal2002tsp}
Peter Kabal.
\newblock {TSP} speech database.
\newblock {\em McGill University, Database Version}, 1(0):09--02, 2002.

\bibitem{OpenAir}
Damian~T Murphy and Simon Shelley.
\newblock {OpenAir}: An interactive auralization web resource and database.
\newblock In {\em Audio Engineering Society Convention 129}. Audio Engineering Society, 2010.

\bibitem{Avad-VR}
David Thery and Brian~FG Katz.
\newblock Anechoic audio and {3D}-video content database of small ensemble performances for virtual concerts.
\newblock In {\em International Congress on Acoustics (ICA)}, 2019.

\bibitem{midi-kunstlerfuge}
Kunstderfuge.
\newblock \url{https://www.kunstderfuge.com/}.
\newblock Accessed: 2025-03-06.

\bibitem{pianobook}
{PianoBook}.
\newblock \url{https://www.pianobook.co.uk/}.
\newblock Accessed: 2025-03-06.

\bibitem{decayfitnet}
Georg G{\"o}tz, Ricardo Falc{\'o}n~P{\'e}rez, Sebastian~J Schlecht, and Ville Pulkki.
\newblock Neural network for multi-exponential sound energy decay analysis.
\newblock {\em The Journal of the Acoustical Society of America}, 152(2):942--953, 2022.

\bibitem{spcmic}
spcmic.
\newblock \url{https://spcmic.com/}.
\newblock Accessed: 2025-03-11.

\bibitem{sadie}
Cal Armstrong, Lewis Thresh, Damian Murphy, and Gavin Kearney.
\newblock A perceptual evaluation of individual and non-individual {HRTFs}: A case study of the {SADIE II} database.
\newblock {\em Applied Sciences}, 8(11):2029, 2018.

\bibitem{peirce2019psychopy2}
Jonathan Peirce, Jeremy~R Gray, Sol Simpson, Michael MacAskill, Richard H{\"o}chenberger, Hiroyuki Sogo, Erik Kastman, and Jonas~Kristoffer Lindel{\o}v.
\newblock Psychopy2: Experiments in behavior made easy.
\newblock {\em Behavior research methods}, 51(1):195--203, 2019.

\bibitem{beyer}
{DT} 990 {PRO}.
\newblock \url{https://europe.beyerdynamic.com/p/dt-990-pro}.
\newblock Accessed: 2025-04-03.

\end{thebibliography}

\end{document}